\documentstyle[prb,preprint,aps,epsf]{revtex}
\begin{document}
\title{ Exchange Narrowing Effects in the EPR Linewidth of $Gd$ Diluted in $Ce$ Compounds}
\draft
\author{Pablo A. Venegas\cite{santacruz} \\} 
\address{Physics Department, University of California \\
Santa Cruz, CA 95064}
\author{Paulo R. S. Netto \\}
\address{Departamento de F\'{\i}sica, Universidade Estadual Paulista \\
Caixa Postal 473, 17033-360, Bauru-SP, Brazil}
\date{\today}
\maketitle
\begin{abstract}

Anomalous thermal behavior on the EPR linewidths of $Gd$ impurities
diluted in $Ce$ compounds has been observed.  In metals, the local
magnetic moment EPR linewidth, $\Delta H$, is expected to increase
linearly with the temperature. In contrast, in $Ce_{x}La_{1-x}Os_{2}$ the
Gd EPR spectra show a nonlinear increase.  In this work, the mechanisms
that are responsible for the thermal behavior of the EPR lines in
$Ce_{x}La_{1-x}Os_{2}$ are examined. We show that the exchange interaction
between the local magnetic moments and the conduction electrons are
responsible for the narrowing of the spectra at low temperatures. At high
temperatures, the contribution to the linewidth of the exchange
interaction between the local magnetic moments and the $Ce$ ions has an
exponential dependence on the excitation energy of the intermediate valent
ions. A complete fitting of the EPR spectra for powdered samples is
obtained. 

\end{abstract}

\section{INTRODUCTION} 

During the last two decades $Ce$ compounds have been extensively studied
and many of them show anomalies typical of intermediate valence systems
\cite{robinson}. Electron paramagnetic resonance (EPR) spectra of magnetic
ions diluted in $Ce$ compounds allow to study locally the influence of the
intermediate valent $Ce$ ions.  In normal metals, the thermal behavior of
the local moments EPR linewidth, $\Delta H$, is linear and described by
the Korringa mechanism \cite{korringa}.  In contrast, in
$Ce_{x}A_{1-x}Pd_3$ ($A = Ag, Y$) and $Ce_{x}La_{1-x}OS_2$ the $Gd$
spectra shows a non-linear increase of $\Delta H$ in the temperature range
$4.2 < T < 300K$, and a strong dependence on the $Ce$
concentration\cite{schaeffer,schlott}.  At low temperatures (T) the slope
$d(\Delta H)/ dT$ is smaller than in the isostructural non-IV compounds
$MPd_3$ ($M = Sc, Y, La$) \cite{gambke1} and $LaOs_2$ 
\cite{schlott}, however, at high temperatures the resonance line is
strongly broadened and the slope asymptotically approaches the value
measured on them.  Then, to describe this abnormal behavior of linewidth
other mechanisms different from the Korringa are necessary. 

It was demonstrated \cite{venegas1}, that at high temperatures, the
indirect exchange interaction between the local magnetic and the $Ce$ $4f$
electrons has an appreciable contribution to the linewidth of the magnetic
impurities at high temperature.  The $Ce$ interconfigurational
fluctuations are transfered via RKKY interaction to the $Gd$ site as an
effective alternating magnetic field, which relaxes the $Gd$ spins. 
However this mechanism is not sufficient to describe the low temperature
behavior of the $Gd$ linewidth. It is well known \cite{urban,venegas2}
 that the $Gd$ EPR spectra shows a resolved fine structure at low T, which
is narrowed due to the exchange interaction between local magnetic moments
and conduction electron when the temperature is increased. Then, for the
correct description of the thermal behavior of the $Gd$ spectra it is
necessary to include not only the intermediate valence effects but also
the narrowing mechanism.  The goal of this work is to calculate the EPR
spectra o $Gd$ diluted in $Ce_xLa_{1-x}Os_2$ including the both mechanisms. 

\section{THEORETICAL ANALYSIS}

To obtain the EPR absorption, the transverse dynamic susceptibility of 
the local magnetic moments coupled to the conduction electrons is necessary. 
The susceptibility, including the crystal field interaction, can be obtained 
using the projector formalism in the Liouville space \cite{plefka}. Normally 
the EPR experiments are performed at concentration that the conduction electrons
 static susceptibility 
 is much smaller than those of the local moments, then, 
in this approximation, the susceptibility  for the system 
in the non-bottleneck regime can be written as follows

\begin{equation}
\chi^{+}(\omega) \approx 1 \, -\, \omega_{0} \left[ \sum_{M,M'} P_{M} 
(\Omega^{-1})_{M,M^{'}}\right]
\end{equation}

\noindent
where $\Omega^{-1}_{M,M^{'}}$ is the transition matrix,  the quantum numbers $M$ and $M'$  
describe the various Zeeman states ($M, M'= -S, -S+ I, . . S - 1$) associated 
to the  $S=7/2$  $Gd$ spin and $P_M$ are 
transition probabilities associated to the $M \leftrightarrow M +1$ transition 
and can be written as:

\begin{equation}
P_{M} = C_{M} \exp ^{M \hbar \omega_{0}/kT} / \sum_{M} C_{M^{'}} 
\exp ^{M^{'} \hbar \omega_{0}/kT}
\end{equation}

\noindent
where $C_M = S(S+1)-M(M+1)$ and $k$ is the Boltzmann constant.  The elements  of the transition matrix for 
$kT$ large compared to $\hbar\omega_0$  are expressed by the formula

\begin{eqnarray}
\Omega_{M,M^{'}}= &(&\hbar \omega_{0} / g \mu_{B} - H - H_{M}) 
\delta_{M,M^{'}} -
i \Delta H_{res} \delta_{M,M^{'}} -  \nonumber \\
&i& \frac{1}{2} b C_{M^{'}} (2 \delta_{M,M^{'}} - \delta_{M,M^{'}}
- \delta_{M,M^{'}+1} - \delta_{M,M^{'}-1}
\end{eqnarray}

\noindent
where $\omega_0$ is the microwave frequency, $H$ the variable external
magnetic field,  $\Delta H_{res}$ is the temperature independent residual
linewidth of the various fine structure lines,
 $b$, the Korringa parameter, $\mu_B$  the Bohr magneton and $H_M$ the resonance field of the  $Gd$ 
$M \rightarrow M +1$ transition. As we can see, the transition matrix 
is a tri-diagonal matrix where the diagonal elements contain the linewidth of each
resonance line and the resonance field. The upper and lower diagonals terms represent 
 the fluctuation rates of the local moment between two consecutives resonance frequencies. 

 In a cubic environment the fine structure spectra is given by 

\begin{eqnarray}
H( \pm \frac{7}{2} \leftrightarrow \pm \frac{5}{2}) &=& H_0 \mp   (1-5 \phi ) b_4 \nonumber \\
H( \pm \frac{5}{2} \leftrightarrow \pm \frac{3}{2}) &=& H_0 \pm   (1-5 \phi ) b_4  \nonumber \\
H( \pm \frac{3}{2} \leftrightarrow \pm \frac{1}{2}) &=& H_0 \pm   (1-5 \phi ) b_4  \nonumber \\
H( +   \frac{1}{2} \leftrightarrow   - \frac{1}{2}) &=& H_0
\end{eqnarray}

\noindent
where $b_4$ is the crystal field parameter for the $Gd$ ion, $\phi$ is given by

\begin{equation}
\phi  =  \sin ^2 \theta \cos ^2 \theta +  \sin ^4 \theta \cos ^2 \varphi \sin ^2 \theta
\end{equation}

\noindent
and $\theta$ and $\varphi$ are spherical coordinates of the applied magnetic 
field $H$ with respect to the axes of the crystal.

As it was showed previously \cite{urban,venegas2}, for the $LaSb:Gd$ and $CePd_3:Gd$, 
 the spin-spin interaction between $Gd$ ions has an important role
in the calculation of the resonance spectra. In both cases was found that 
when the spin-spin interaction is not taken into account the narrowing in the theoretical
 spectra   
 occurs at  higher
temperatures than in the experiment, the transition $1/2 \leftrightarrow 1/2$
does not appear in the theoretical spectra in the intermediate temperature
range,  however, it appears in the experiment, and the experimental
linewidth of the single line at high temperatures is smaller than the
calculated one. The discrepancies between theory and experiment can be
overcome by introducing this interaction.  Unfortunately, no theoretical calculation which
takes spin-spin interaction into account exists at present. We shall 
introduce the spin-spin interaction in
a phenomenological way \cite{urban,venegas2}. This can be done by adding to the 
transition matrix elements the term:

\begin{equation}
\Omega^{ex}_{M,M'} = i \frac{H_{ex}}{P_{M}} \left( 1 - \delta_{M,M^{'}} \right) 
- i \frac{6H_{ex}}{P_{M}} \delta_{M,M'}
\end{equation}

\noindent
where $H_{ex}$ is the exchange-field parameter. The transition matrix has some 
properties that must be satisfied for this additional term. Firstly, 
$P_M$ $Im(\Omega_{M,M'})$ are the elements of a
negative-definite symmetric matrix.  This guarantees a positive energy
absorption.  Secondly, as the total spin commutes with the exchange
Hamiltonian, the relation $\sum_{M} \Omega_{M,M'}=0$ must hold. Since these two requirements are
satisfied, we can believe that the additional term for the transition matrix describes the main
effects of the spin-spin interaction.  Because of the random distribution of the $Gd$
ions, it is realistic to assume a distribution of $H_{ex}$. 
For the present calculations we use  an  slightly modified Lorentzian 
distribution for the exchange-field  with a maximum at
$H_{ex} = 0$. The mean exchange field amounts to about $95\,\, G$ and 
 the distribution function was cut off at $1500\,\, G$.

\section{THE LINEWIDTH}

In our metallic host the impurity EPR linewidth calculations must consider
the energy transfer between the impurity spin S, the spin of the host rare
earth ions ($Ce$), the spins of the conduction electrons as well as the
lattice, in the presence of static external magnetic field and a small
alternating field.  In the present work we shall assume that the system is
in the unbottlenecked regime i.e. the conduction electrons and the host
magnetic ions are in equilibrium with the lattice.  In addition we shall
 assume that the transverse susceptibility associated with the impurity at
the impurity resonance frequency, is much larger than those of the
other spin systems at the same frequency.  In this limit one can consider
only the transverse magnetization, $M_{x}$, of the impurities and neglect the
interaction of the $rf$ field with the magnetic host ions and the conduction
electrons.  In other words, the magnetic host ions and the conduction
electrons are ``passive dissipative systems''.  With these assumptions the
 impurity linewidth can be expressed as:

\begin{equation}
\Delta H = \Delta H_{res}+ bT + \Delta_{IF}
\end{equation}

\noindent
where $\Delta H_{res}$ is the residual linewidth, $b$  the usual Korringa
contribution, originated from the impurity-conduction-electron exchange
interaction and $\Delta_{IF}$ is the impurity linewidth due to the exchange
interaction with the $Ce$ host ions \cite{venegas1}.  The last contribution arises from the
RKKY coupling between the $Ce$ and $Gd$ ions, which transfers the $Ce$
fluctuations to the $Gd$ site. $Ce$ ions, fluctuate between the $4f^{0}$ and the $4f^{ 1}$
configurations.  If we assume the $4f^{ 0}$ configuration as the ground state
(with $J_z = 0$), the contribution of the excited $4f^1$ configuration to the
impurity linewidth can be written as: 

\begin{equation}
 \Delta_{IF} = A e^{-E_{ex}/T}
\end{equation}

\noindent
where $E_{ex}$ is the energy required to delocalize a single
electron from the $Ce$ $4f^1$ configuration and $A$ is an adjustable parameter
defined in Ref. [\onlinecite{venegas1}]. 

The resonance absorption $P$ is calculated using the relation 
$P = [Re(\chi^+(\omega)) - Im(\chi^+(\omega))]$.
To obtain the EPR linewidth of the powdered samples we have to
integrate the absorption over all directions of the magnetic field.

\section{RESULTS AND DISCUSSION}

 Using the results above we calculate the total linewidth of
the $Gd$ resonance.  Figure \ref{fig1} shows the calculated and 
experimental linewidth of $Ce_xLa_{1-x}Os_2$ powdered sample, as a
function of  temperature for selected concentrations of $Ce$.  The residual linewidth is sample
dependent and it was not considered in the plot. To fit the data 
we adjust the parameters $b$, $b_4$, $A$ and $E_{ex}$ using the 
Monte Carlo - Simulated Annealing. With this technique  
is possible to obtain more accurate results than previous 
calculations  with conventional methods \cite{venegas1,urban,venegas2}. Comparing
our theoretical results with the experimental data we can see that in all
of the cases we got an excellent fit.  According to our model, at low T
the main contribution to the linewidth is originated in the usual Korringa
mechanism and the exchange narrowing effects.  At high temperatures the population of the excited $Ce$ $4f^1$
configuration is increased, and the exponential contribution given by $\Delta_{IF}$
is the most important.  The parameters used to fit the experimental data
are shown in Table 1.
The analysis {\it a priori} of the exchange narrowing effects in this system is not 
easy  because there do not exist a single crystal spectra to  obtain by a direct measure 
the crystal field parameter, however if we look for the spectra of
$Ce_{0.8}La_{0.2}Os_2$  in Fig. 1a of Ref.[\onlinecite{schlott}]  
we can see clearly that the Dysonian type line does not fit 
the experimental result. The experimental  spectra shows the typical broadening at low $T$ due 
to the fine structure contribution, and these effects tend to be more important in higher Ce 
concentrations \cite{venegas2}. If we compare the exchange narrowing 
effects for the present case with that  in $CePd_3$ \cite{venegas2}
certainly the exchange narrowing effects are less important in the former one, due to the smaller 
crystal field parameter. Otherwise, its contribution is more important at lower temperatures 
than in $CePd_3$ because  the higher Korringa parameter values collapses the spectra at lower temperatures.
 It is necessary to point out that the minimization results show clearly 
a non zero value for the crystal field parameter which reflects
the importance of this mechanism.  
On the other hand, looking for the intermediate valence mechanism,  the effect of the
Ce concentration at high temperature seems clear.  The increase in the excitation energy value
when the $Ce$ concentration is reduced agree with the interconfigurational
fluctuation model \cite{hirst1,hirst2,hirst3,gambke2}.  This model predicts an increasing on the
excitation energy when we cross from the intermediate valence to the
magnetic regime.  On the other hand, the $A$ value depends on the strength
of the $Ce$ fluctuation spectra and we can expect an increase with the $Ce$
concentration.  The small values of $b$ obtained for high Ce concentrations 
(when compared with those of  $LaOs_2$)
agree with the experimental results of Ref. [\onlinecite{schlott}] at low $T$, with 
 the prediction of the ``hybridization hole'' model \cite{schaeffer}
for $Gd$ diluted in $CePd_3$ powdered sample and with that obtained in Ref.[{\onlinecite{venegas2}]
for the monocristalline spectra of the same compound.  This result agree
also with that obtained by Hirst \cite{hirst3}, for low T, but not with the result
obtained in Ref. [\onlinecite{ochi}], which predicts a higher value for $b$. However, the
thermal behavior of the linewidth obtained by the latter authors agree, at
least qualitatively, with that obtained here. It is important to observe
that in contrast to the hybridization hole model, here we have supposed a
constant density of states as in a normal metal.  The non linear
contribution to the linewidth, at high $T$, according with the present model, is
originated in the exchange interaction of the magnetic impurities with
the $Ce$ ions.  

Concluding, our calculations  using the intermediate valence and exchange 
narrowing mechanism  permit a quantitative description of the 
thermal behavior of $Gd$
linewidths. Note that the values obtained for the excitation energy 
are, within the experimental error, close to that founded by 
Sereni et al. \cite{sereni}.  The agreement with experimental 
results supports our interpretation.

\acknowledgments PAV thanks the Brazilian Agency Capes for partial 
financial support and PRSN thanks the Brazilian agency CNPQ
for financial support.

\begin{figure} 
\epsfxsize=1.1\columnwidth
\epsfysize=1.1\columnwidth
\centerline{\epsffile{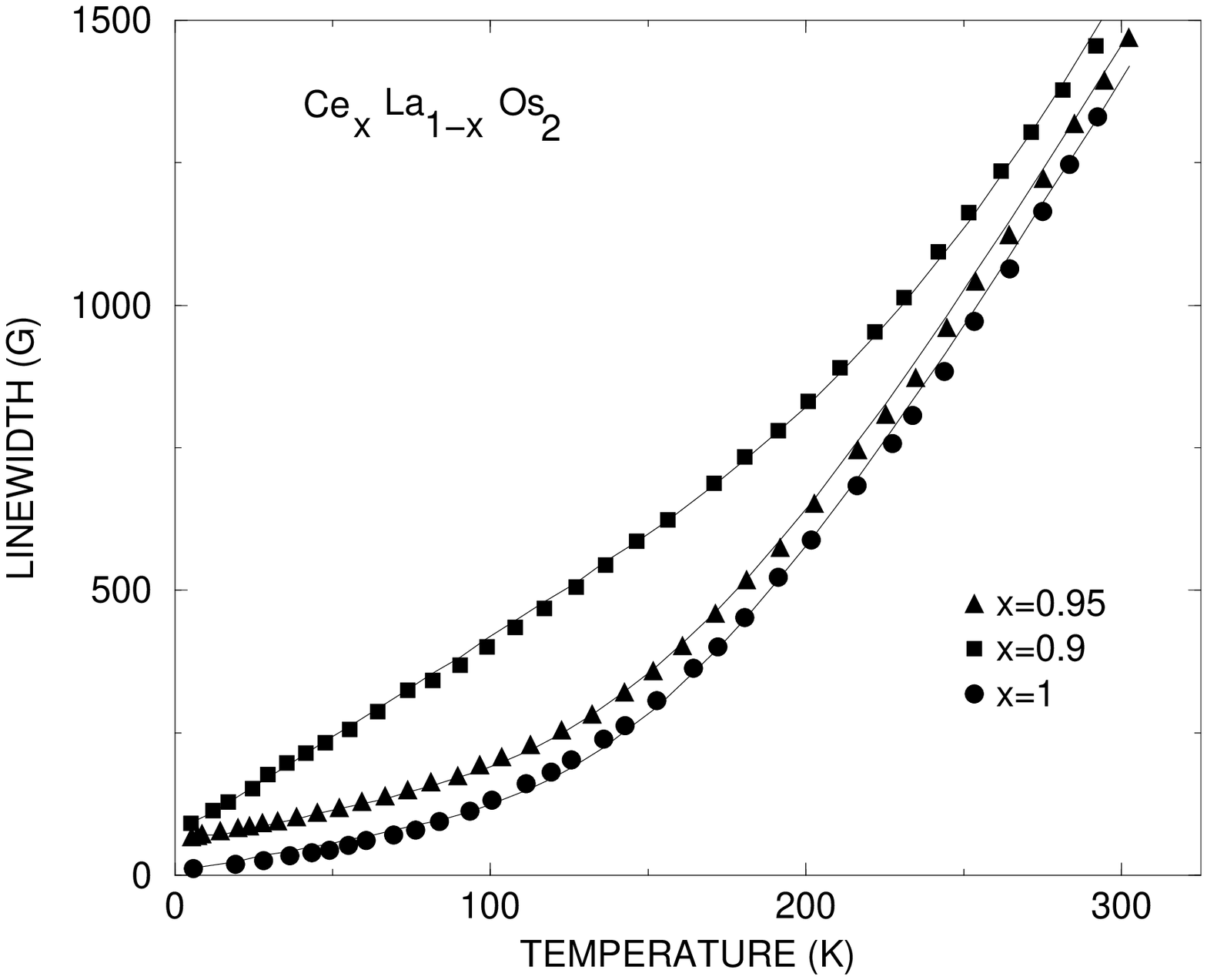}} 
\vspace{0.1cm} \caption{Temperature
dependence of the $Gd$ linewidth for selected values of concentration ($x$)
in $Ce_xLa_{1-x}Os_2$. The full lines represent the theoretical results.
The residual linewidths are not realistic and the data has been shifted to
avoid curve overlap.} 
\label{fig1} 
\end{figure}

\begin{table}
\caption{Obtained  parameters $b$, $b_4$, $A$, $E_{ex}$ that fit the 
the linewidth thermal behavior for $Gd$ diluted in $Ce_xLa_{1-x}Os_2$. 
$E_{ex}$ (expt.) is the experimental value of $E_{ex}$ to be taken as a comparison.
}
\begin{tabular}{lccccc}

$x$   & $b[G/K]$ &  $b_4 [G]$  &  $A [G]$  &  $E_{ex}$ (theor.)  &  $E_{ex}$ (expt.)  \\
\hline 
1.0 &  1.07  & 4.8        & 10200   &   670               &     500		\\
0.95&  1.32  & 10.6       & 10150   &   684               &\\
0.9 &  3.5   & 9.0        & 9800   &   1243              &  \\
\end{tabular}
\end{table}

\end{document}